\documentclass[12pt]{article}
\usepackage{amssymb,latexsym}
\usepackage[dvips]{graphicx}
\def\bea{\begin{eqnarray}}
\def\eea{\end{eqnarray}}
\def\nn{\nonumber}
\def\beq{\begin{equation}}
\def\eeq{\end{equation}}

\def\nn{\nonumber}

\def\ra{\rangle}
\def\la{\langle}
\def\lb{[\![}
\def\rb{]\!]}
\def\l{\ldots}
\def\t{\theta}

\begin{document}
%
%
\begin{center}
{\Large \bf
Microscopic and macroscopic properties\\
of $A$-superstatistics}\\[5mm]
{\bf T.D.\ Palev}\footnote{E-mail~: tpalev@inrne.bas.bg.}\\[1mm]
Institute for Nuclear Research and Nuclear Energy,\\
Boul.\ Tsarigradsko Chaussee 72, 1784 Sofia, Bulgaria;\\[2mm]
{\bf N.I.\ Stoilova}\footnote{E-mail~: Neli.Stoilova@UGent.be.
Permanent address~: Institute for Nuclear Research and Nuclear Energy,
Boul.\ Tsarigradsko Chaussee 72, 1784 Sofia, Bulgaria.}
{\bf and J.\ Van der Jeugt}\footnote{E-mail~:
Joris.VanderJeugt@UGent.be.}\\[1mm]
Department of Applied Mathematics and Computer Science,
University of Ghent,\\
Krijgslaan 281-S9, B-9000 Gent, Belgium.
\end{center}

\vskip 15mm

\vskip 10mm
\begin{abstract}\noindent
The microscopic and the macroscopic properties of $A$-superstatistics, 
related to the class $A(0,n-1)\equiv sl(1|n)$ of simple Lie superalgebras
are investigated. The algebra $sl(1|n)$ is described in terms
of generators $f_1^\pm, \ldots, f_n^\pm$, which satisfy certain triple
relations and are called Jacobson generators. The Fock spaces of 
$A$-superstatistics are investigated and the Pauli principle of the corresponding
statistics is formulated. Some thermal properties of $A$-superstatistics
are constructed under the assumption that the particles interact only 
via statistical interaction imposed by the Pauli principle.
The grand partition function and the average number of particles are written
down explicitly in the general case and in two particular examples~: 1) the particles
have one and the same energy and chemical potential; 2) the energy spectrum
of the orbitals is equidistant.
\end{abstract}
\vfill\eject

\section{Introduction}
\setcounter{equation}{0}

In this paper we consider the microscopic and the macroscopic properties
of a class of generalized statistics, referred to as $A$-superstatistics.
In our approach, the starting point is a certain symmetry principle,
characterized by an algebra of creation and annihilation operators (CAO's)
with Fock type representations.

The idea behind these investigations is based on a few observations. The first one is
that any $n$ pairs of Bose CAO's $B_1^\pm, B_2^\pm, \ldots , B_n^\pm$ generate a
representation, the Bose representation $\rho _B$, of the orthosymplectic Lie
superalgebra $osp(1|2n)=B(0,n)$. The representation independent generators
$\hat{B}_1^\pm, \hat{B}_2^\pm, \ldots , \hat{B}_n^\pm$ of $osp(1|2n)$, which in the
Bose representation coincide with the Bose operators $\rho_B(\hat{B}_i^\pm )= B_i^\pm$,
are para-Bose CAO's~\cite{Green} and satisfy the relations
\begin{equation}
[\{\hat{B}_i^\xi,\hat{B}_j^\eta\},\hat{B}_k^\epsilon]=
(\epsilon-\xi)\delta_{ik}\hat{B}_j^\eta
+(\epsilon-\eta)\delta_{jk}\hat{B}_i^\xi, \quad
\xi, \eta, \epsilon =\pm\hbox{ or }\pm 1.
\label{pB}
\end{equation}
These triple relations give one possible definition of the Lie superalgebra
$osp(1|2n)$~\cite{OmoteGanchev}, and are considered as the defining relations of
para-Bose statistics. So ordinary Bose statistics corresponds to
one particular realization of para-Bose statistics.

The situation with Fermi statistics and its generalization, para-Fermi
statistics~\cite{Green}, is
similar.
So the second observation is that any $n$ pairs of Fermi CAO's $F_1^\pm,
F_2^\pm, \ldots , F_n^\pm$ give a representation, the Fermi representation $\rho_F$, of
the orthogonal Lie algebra $so(2n+1)=B_n$ and any $n$ pairs of para-Fermi CAOs
$\hat{F}_i^\pm $, with
\begin{equation}
[[\hat{F}_i^\xi,\hat{F}_j^\eta],\hat{F}_k^\epsilon]= {1\over
2}(\eta-\epsilon)^2\delta_{jk}\hat{F}_i^\xi -{1\over
2}(\xi-\epsilon)^2\delta_{ik}\hat{F}_j^\eta,
\quad\xi, \eta, \epsilon =\pm\hbox{ or }\pm 1,
\label{pF}
\end{equation}
generate the algebra $B_n$~\cite{KR}.

Both algebras $B_n$ (every Lie algebra is also a Lie superalgebra) and $B(0|n)$ belong
to the class $B$ of basic classical Lie superalgebras in the classification of
Kac~\cite{Kac}. Hence ordinary Bose and Fermi statistics and their generalizations
para-Bose and para-Fermi statistics could be referred to as $B$-(super)statistics.

On the ground of these facts statistics related to the other classes of basic Lie
superalgebras were introduced~: $A$-, $B$-, $C$- and
$D$-(super)statistics~\cite{Palev1}. So far only $A$-statistics (corresponding to the
Lie algebra $sl(n+1)$) was studied in detail from the
microscopic~\cite{Palev1},~\cite{PalevJeugt} and macroscopic point of
view~\cite{Jellal}. In the present paper we investigate the properties of
$A$-superstatistics~\cite{Palev2}, namely the statistics arising
from the Lie superalgebra $sl(1|n)=A(0,n-1)$ or $gl(1|n)$.

The microscopic properties of this statistics related to $gl(1|n)$ lead
to the introduction of Fermi-like operators $f_i^\pm$, and these are relevant 
for the description of certain physical models (see Section 3). 
In particular, lattice models  of strongly correlated electron systems use 
these operators $f_i^\pm$ (in a 
certain representation). Also, the so-called ideal odd-particle creation and
annihilation operators of Klein and Marshalek in nuclear shell model theory 
correspond to a particular realization of the operators $f_i^\pm$.

Section 2 is devoted to the microscopic properties of this new class
of generalized statistics. First we recall the definitions
of the Lie superalgebras $sl(1|n)$ and $gl(1|n)$ and their Fock
representations. Like for para-Bose and para-Fermi statistics,
the generators $f_i^\pm$, $i=1,\ldots ,n$ of $sl(1|n)$ satisfy certain
triple relations. These relations completely define the Lie superalgebra $sl(1|n)$,
just as the triple relations for the generators $a_i^\pm$,
$i=1,\ldots , n$ of $A$-statistics do for the Lie algebra $sl(n+1)$~\cite{PalevJeugt}.
For $sl(n+1)$, this property of the operators $a_i^\pm$ was first
observed by Jacobson~\cite{Jacobson} and therefore the generators $a_i^\pm$ are
referred as Jacobson generators of $sl(n+1)$~\cite{PalevJeugt}.
By analogy we call the generators $f_i^\pm$ of $sl(1|n)$
also Jacobson generators of $sl(1|n)$.
The Fock representations of $A$-superstatistics are constructed
in the same way as in parastatistics. They are generated by the
operators $f_i^+$ and labeled by a positive integer $p=1,2,\ldots .$
Within the corresponding module $W(p,n)$ the generator $f_i^+$
(respectively $f_i^-$) can be interpreted as a creation (respectively
annihilation) operator of a ``particle" on the $i$th orbital ($i=1,2,
\ldots ,n$). The Pauli principle of $A$-superstatistics is formulated
and in Section 3 it is indicated that in the limit $p \rightarrow \infty$
the representation dependent operators $F(p)_i^\pm =
{f_i^\pm\over \sqrt {p}}$ coincide with ordinary Fermi CAO's.
We complete this section by showing that the Jacobson generators of
$sl(1|n)$ are implicitly present in certain physical models.

The following sections are devoted to the macroscopic properties of
$A$-super\-statis\-tics. In Section 4 we construct explicitly the $sl(1|n)$
grand partition function $Z(p,n)$ (GPF), the average number of particles
in the system $\bar{N}(p,n)$ and  the average number of particles on each
orbital $\bar{\t}_i$  under
the assumption that the energy of each particle on orbital $i$ is $\epsilon_i$.
All these thermal properties of the system are described by means of
the elementary symmetric functions.
The fact that symmetric functions appear naturally in the description
of (grand) partition functions in statistical mechanics got
attention recently~\cite{Schmidt}. For quantum systems with Bose
or Fermi statistics, see~\cite{Schmidt}. For quantum systems with
$A$-statistics, the relevant symmetric functions are the
so-called complete symmetric functions~\cite{Jellal}.
Here in the case of $A$-superstatistics, the relevant symmetric
functions are the so-called elementary symmetric functions.

Then we consider two specializations of the general case of Section 4.
In Section 5 the energy and the chemical potential of each orbital
are assumed to be the same. This is the so-called degenerate case.
The thermodynamical functions simplify, and many of these can be
expressed as hypergeometric series. Furthermore, these functions
can be seen as deformations of the corresponding ones in the case
of Fermi statistics.
In Section 6 we investigate a model with equidistant energy levels.
Also under this specialization, the thermodynamical functions assume
a simple form, usually in terms of $q$-generalized or basic hypergeometric
functions.

It should be emphasized that our notion ``$A$-superstatistics'' refers to 
a generalized quantum statistics based upon the Lie superalgebra of type $A$, 
and that it should not be confused with the recently introduce phenomenon of 
``superstatistics'' as in~\cite{Beck}.

\section{Microscopic properties of $A$-su\-per\-sta\-tistics}
\setcounter{equation}{0}

$A$-superstatistics is defined in the context of the Lie superalgebra $sl(1|n)$ or
$gl(1|n)$. A  convenient 
basis of $gl(1|n)$ is given with the
 Weyl generators 
$e_{ij}$, with $i,j\in\{0,1,\ldots,n\}$. The grading of $gl(1|n)$ is as follows~: the
{\em even} elements are given by $e_{00}$ and $e_{ij}$ with $i,j\in\{1,\ldots,n\}$; the
{\em odd} elements are $e_{0i}$ and $e_{i0}$ ($i=1,\ldots,n$). The Lie superalgebra
bracket is determined by
\begin{equation}
\lb e_{ij}, e_{kl} \rb \equiv e_{ij} e_{kl}-(-1)^{{\deg(e_{ij})
\deg(e_{kl})}}e_{kl}e_{ij} = \delta_{jk} e_{il} - (-1)^{\deg(e_{ij}) \deg(e_{kl})}
\delta_{il} e_{kj}, \label{Weyl}
\end{equation}
where $\deg(e_{ij})$ is 0 (resp.\ 1) if $e_{ij}$ is even (resp.\ odd).
One can define $sl(1|n)$ as the (super)commutator algebra of $gl(1|n)$;
its basis consists of all elements $e_{ij}$ ($i\ne j$) and
the Cartan elements $e_{00}+e_{ii}$ ($i=1,\ldots,n$).

The Jacobson creation and annihilation operators $f_i^\pm$ of $sl(1|n)$
are given by
\begin{equation}
f_i^+ = e_{i0}, \qquad f_i^- = e_{0i},\qquad (i=1,\ldots,n).
\end{equation}
It is known~\cite{Palev2}  that the linear envelope of
\begin{equation}
\left\{ f_i^\xi, \{ f_j^\eta, f_k^\epsilon\} | i,j,k=1,\ldots,n;
\xi,\eta,\epsilon=\pm \right\}
\end{equation}
is indeed the Lie superalgebra $sl(1|n)$.

$A$-superstatistics is determined by the relations that hold
for the creation and annihilation operators. These are~:
\begin{eqnarray}
&& \{ f_i^+, f_j^+ \} = \{ f_i^-, f_j^- \} = 0 , \label{2.4} \\
&& [ \{ f_i^+, f_j^- \}, f_k^+ ] = \delta_{jk} f_i^+ - \delta_{ij} f_k^+,\\
&& [ \{ f_i^+, f_j^- \}, f_k^- ] = -\delta_{ik} f_j^- + \delta_{ij} f_k^- . \label{2.6}
\end{eqnarray}
It is worth observing that these operators satisfy the compatibility
conditions required in the context of Wigner quantum systems~\cite{WQS}.

The Fock representations of $A$-superstatistics have been classified
by Palev~\cite{Palev2}. These representations are labeled by a positive integer $p$,
the order of statistics.
Let us denote the Fock representation of order $p$ for $sl(1|n)$
by $W(p,n)$. The space $W(p,n)$ is characterized by a vacuum
vector $| 0 \ra$, such that
\begin{eqnarray}
&& f_i^- |0\ra = 0 , \qquad (i=1,\ldots,n),\\
&& f_i^- f_j^+ |0\ra = p\; \delta_{ij} |0\ra,\qquad (i,j= 1,\ldots,n).
\end{eqnarray}
These Fock spaces are finite-dimensional unitary irreducible
$sl(1|n)$-modules. A set of basis vectors for the space $W(p,n)$
consists of all vectors
\begin{equation}
(f_1^+)^{\t_1} (f_2^+)^{\t_2} \ldots (f_n^+)^{\t_n} |0\ra,
\qquad \t_i\in\{ 0,1\}
\label{allvectors}
\end{equation}
where
\begin{equation}
|\t|\equiv \sum_{i=1}^n \t_i \leq p.
\label{restriction}
\end{equation}
The linear  span
of all vectors (\ref{allvectors}), without the restriction~(\ref{restriction}),
also forms a $sl(1|n)$ module $\bar W(p,n)$. However, if $p<n$, $\bar W(p,n)$ is not
irreducible~: it contains a maximal invariant submodule, and $W(p,n)$ is the quotient
module of $\bar W(p,n)$ with respect to this submodule. If $p\geq n$, we have that
$\bar W(p,n)=W(p,n)$; in this case it is clear that the restriction~(\ref{restriction})
is superfluous.

One can define a Hermitian form $\langle\,,\,\rangle$ on $W(p,n)$ with the
usual Fock space technique, by requiring
\begin{eqnarray}
&& \la 0| 0\ra  =1,\\
&& \la f_i^\pm v | w \ra = \la v | f_i^\mp w \ra, \qquad\forall v,w\in W(p,n).
\end{eqnarray}
With respect to this form, the different vectors in (\ref{allvectors})
are orthogonal, and the following vectors form an orthonormal
basis of $W(p,n)$~:
\begin{eqnarray}
&& |p;\t\ra \equiv |p;\t_1,\ldots,\t_n\ra =
\sqrt{(p-|\t|)!\over p!}(f_1^+)^{\t_1} (f_2^+)^{\t_2}
\ldots (f_n^+)^{\t_n} |0\ra, \nn\\
&&\qquad \t_i\in\{ 0,1\},\quad  |\t|\leq p.
\label{orthvectors}
\end{eqnarray}
Furthermore, the Hermitian conjugate of $f_i^\pm$ is $f_i^\mp$ in this
module, which is an important physical requirement.

The transformation of the basis (\ref{orthvectors}) under the action
of the creation and annihilation operators reads~:
\begin{eqnarray}
&& f_i^- |p;\t\ra = \t_i (-1)^{\t_1+\cdots+\t_{i-1}}
\sqrt{p-|\t|+1}\; |p;\t_1,\ldots,\t_i-1,\ldots,\t_n\ra, \label{fi-}\\
&& f_i^+ |p;\t\ra = (1-\t_i) (-1)^{\t_1+\cdots+\t_{i-1}}
\sqrt{p-|\t|} \; |p;\t_1,\ldots,\t_i+1,\ldots,\t_n\ra. \label{fi+}
\end{eqnarray}
Our Hamiltonian will be an element from (the Cartan subalgebra of) $gl(1|n)$. Therefore
we first
extend $W(p,n)$ to a $gl(1|n)$ module. For this
purpose, we set $N_i=e_{ii}$ ($i=0,1,\ldots,n$), and have~:
\begin{eqnarray}
&& N_0 |p;\t\ra = (p-|\t|) |p;\t\ra, \label{N0}\\
&& N_i |p;\t\ra = \t_i |p;\t\ra. \label{Ni}
\end{eqnarray}

We shall be studying macroscopic properties of $A$-superstatistics
for a Hamiltonian of the following form~:
\begin{equation}
H = \sum_{i=1}^n \epsilon_i N_i.
\label{H}
\end{equation}
Via creation and annihilation operators, this can be rewritten as
\begin{equation}
H = \sum_{i=1}^n \epsilon_i \left( \{f_i^+,f_i^-\} +{1\over{n-1}} \Big( p - \sum_{k=1}^n
\{f_k^+,f_k^-\}\Big) \right). \label{Hf}
\end{equation}
Clearly, $H|0\ra =0$ (so the vacuum has zero energy), and
\begin{equation}
[H,f^\pm_i] = \pm\epsilon_i f_i^\pm.
\end{equation}
Therefore, each $f_i^+$ (resp.\ $f_i^-$) can be interpreted as
an operator creating (resp.\ annihilating) a particle
(or quasiparticle, or excitation) on orbital~$i$ (with
energy~$\epsilon_i$). Since
\begin{equation}
H |p;\t\ra = \left(\sum_{i=1}^n \epsilon_i \t_i\right) |p;\t\ra,
\end{equation}
$|p;\t\ra$ is interpreted as a state with $\t_i$ particles on
orbital~$i$.

The CAO's of $sl(1|n)$ together with the Hamiltonian~(\ref{H}) generate $gl(1|n)$. Then,
in view of (\ref{Hf}), $f_1^\pm,\ldots,f_n^\pm$, considered as operators in any
$W(p,n)$, generate $gl(1|n)$. For this reason we call the CAO's of $sl(1|n)$ also
Jacobson generators~\cite{PalevJeugt} of $gl(1|n)$.

The {\em Pauli principle} for $A$-superstatistics follows
basically from equation~(\ref{restriction}) and the form
of the vectors~(\ref{orthvectors}). These relations imply that
the system can accommodate up to $\min(p,n)$,
but no more than $\min(p,n)$
particles, in such a way that every orbital contains at
most one particle.

Note that in the atypical cases $(p<n)$ the Pauli principle introduces an interaction,
a statistical interaction, between the orbitals. For instance if the system is in the 
state with $\t_1=\ldots=\t_p=1$
and $\t_{p+1}=\ldots=\t_n=0$, then it cannot accommodate  more particles, not  even
on the empty orbitals  with $i>p$. Therefore the orbitals are not filled 
independently. The filling of a given orbital depends on fillings of all other orbitals,
which is the main feature of exclusion statistics~\cite{Haldane}. Observe also that
$A$-superstatistics
is closely related to ordinary Fermi statistics~: the only extra condition comes from
the order of statistics~$p$.

\section{Quasi-Fermi creation and annihilation operators}
\setcounter{equation}{0}

In the present section we
discuss some differences and similarities between
$A$-super\-statis\-tics and ordinary Fermi statistics.
First of all, observe that there is an important difference between
the Fock spaces $W(p,n)$ with $p<n$ and those with $p\geq n$.
For $p<n$, the irreducible $sl(1|n)$ representations are {\em atypical},
and their dimension is given by~:
\begin{equation}
    \dim W(p,n) = \sum_{k=0}^p {n \choose k}.
    \label{dim-atypical}
\end{equation}
For $p\geq n$, the representations $W(p,n)$ are all typical, with
\begin{equation}
    \dim W(p,n)=2^n.
    \label{dim-typical}
\end{equation}
For $p_1\ne p_2$ and $p_1,p_2\geq n$, the representations
$W(p_1,n)$ and $W(p_2,n)$ are certainly not isomorphic (even though
they have the same dimension), since they have a different highest
weight.

Let $n$ be fixed, $p\geq n$, and let us define representation-dependent
operators $F(p)_i^\pm$ in $W(p,n)$ by
\begin{equation}
F(p)_i^\pm = {f^\pm_i \over \sqrt{p}}, \qquad i=1,2,\ldots,n.
\end{equation}
The action of these new creation and annihilation operators
on the vectors~(\ref{orthvectors}) reads~:
\begin{eqnarray}
F(p)_i^- |p;\t\ra &=& \t_i (-1)^{\t_1+\cdots+\t_{i-1}}
\sqrt{1+{1-|\t|\over p}} \; |p;\t_1,\ldots,\t_i-1,\ldots,\t_n\ra, \\
F(p)_i^+ |p;\t\ra &=& (1-\t_i) (-1)^{\t_1+\cdots+\t_{i-1}}
\sqrt{1-{|\t|\over p}} \; |p;\t_1,\ldots,\t_i+1,\ldots,\t_n\ra.
\end{eqnarray}
On the other hand, one can consider a set of $n$ ordinary Fermi
creation and annihilation operators $F_i^\pm$ ($i=1,\ldots,n$),
satisfying~:
\begin{equation}
\{ F_i^+,F_j^+\} = \{ F_i^-,F_j^-\} = 0, \qquad
\{ F_i^-,F_j^+\} = \delta_{ij},
\label{Fermi-anticomm}
\end{equation}
and its $2^n$-dimensional Fock space $W(n)$ with orthonormal
basis vectors
\begin{equation}
|\t\ra = |\t_1,\ldots,\t_n\ra=
(F_1^+)^{\t_1} \cdots (F_n^+)^{\t_n} |0\ra,
\qquad \t_i\in\{ 0,1\} ,
\end{equation}
and action
\begin{eqnarray}
 F_i^- |\t\ra &=& \t_i (-1)^{\t_1+\cdots+\t_{i-1}}
 |\t_1,\ldots,\t_i-1,\ldots,\t_n\ra, \label{Fi-}\\
 F_i^+ |\t\ra &=& (1-\t_i) (-1)^{\t_1+\cdots+\t_{i-1}}
|\t_1,\ldots,\t_i+1,\ldots,\t_n\ra. \label{Fi+}
\end{eqnarray}
We can now identify the basis vectors $|p;\t\ra$ of $W(p,n)$
with the vectors $|\t\ra$ of $W(n)$.
Let us therefore define, for a given positive integer $p\geq n$, in $W(n)$ a
realization of the operators $F(p)_i^\pm$,
denoted by $\rho(F(p)_i^\pm)$, through the action~:
\begin{eqnarray}
 \rho(F(p)_i^-) |\t\ra &=& \t_i (-1)^{\t_1+\cdots+\t_{i-1}}
\sqrt{1+{1-|\t|\over p}} \; |\t_1,\ldots,\t_i-1,\ldots,\t_n\ra, \label{3.39}\\
 \rho(F(p)_i^+) |\t\ra &=& (1-\t_i) (-1)^{\t_1+\cdots+\t_{i-1}}
\sqrt{1-{|\t|\over p}} \; |\t_1,\ldots,\t_i+1,\ldots,\t_n\ra. \label{3.40}
\end{eqnarray}
Both $\rho(F(p)_i^\pm)$ and $F_i^\pm$ are now operators acting
in the finite-dimensional Fock space $W(n)$, and from their action
it follows immediately that
\begin{equation}
\lim_{p\rightarrow\infty} \rho(F(p)_i^\pm) = F_i^\pm.
\end{equation}
For this reason, the operators $F(p)_i^\pm$ are said to be
quasi-Fermi creation and annihilation operators.
For large $p$-values, they tend to ordinary Fermi creation
and annihilation operators.

It is interesting to consider the anti-commutators of the quasi-Fermi
creation and annihilation operators, acting in the Fermi Fock space $W(n)$.
{}From~(\ref{3.39})-(\ref{3.40}) one obtains
\begin{eqnarray}
 &&\{ \rho(F(p)_i^+), \rho(F(p)_j^+)\} |\t\ra = 0,\\
 &&\{ \rho(F(p)_i^-), \rho(F(p)_j^-)\} |\t\ra = 0,\\
 &&\{ \rho(F(p)_i^-), \rho(F(p)_i^+)\} |\t\ra =
  \Big( 1+{\t_i-|\t |\over p}\Big)|\t\ra,\\
 &&\{ \rho(F(p)_i^-), \rho(F(p)_j^+)\} |\t\ra =
 -{1\over p}(-1)^{\t_i+\ldots +\t_j}\t_i (1-\t_j) \nn \\
 && \qquad\qquad\qquad\qquad\qquad\times
 |\ldots,\t_i-1,\ldots,\t_j+1,\ldots \ra, \quad i<j,\\
 &&\{ \rho(F(p)_i^-), \rho(F(p)_j^+)\} |\t\ra =
 -{1\over p}(-1)^{\t_j+\ldots +\t_i}\t_i (1-\t_j) \nn \\
 && \qquad\qquad\qquad\qquad\qquad\times
 |\ldots,\t_j+1,\ldots,\t_i-1,\ldots \ra, \quad i>j.
\end{eqnarray}
Compare again with~(\ref{Fermi-anticomm}) to see that the above
anti-commutators tend to ordinary Fermi anticommutators when
$p$ tends to infinity. Also in this sense, the quasi-Fermi creation and
annihilation operators $\rho(F(p)_i^\pm)$ can be considered
as ``deformations'' of ordinary Fermi creation and annihilation
operators, with the integer $p$ (the order of statistics) as
a deformation parameter.

A comparison of~(\ref{fi-})-(\ref{fi+}) for $p\geq n$
with~(\ref{Fi-})-(\ref{Fi+})  shows that the Jacobson
generators of $sl(1|n)$ can be expressed as functions of ordinary Fermi
creation and annihilation operators $F_1^\pm, F_2^\pm, \ldots ,
F_n^\pm$.
Let $W(p,n)$ be a typical representation (so with $p\geq n$), and
identify again its basis vectors $|p;\t_1,\ldots,\t_n\ra$ with the
basis vectors $|\t_1,\ldots , \t_n\ra \equiv|\t \ra $ of the Fermi
Fock space $W(n)$. Since $F_i^+F_i^-$ is a number operator for
fermions in a state $i$,
\begin{equation}
F_i^+F_i^-|\t \ra =\t_i|\t \ra, \qquad i=1,\ldots ,n,
\end{equation}
we can write for~(\ref{fi+})
\begin{equation}
f_i^+|\t \ra =e_{i0}|\t \ra=
(1-\t_i) (-1)^{\t_1+\cdots+\t_{i-1}}
\sqrt{p+1-\sum_{k=1}^n F_k^+F_k^-}\;|\t_1,\ldots ,\t_i+1,\ldots , \t_n\ra.
\end{equation}
The latter can be represented as
\begin{equation}
f_i^+|\t \ra =
\sqrt{p+1-\sum_{k=1}^n F_k^+F_k^-}\;F_i^+|\t\ra =F_i^+
\sqrt{p-\sum_{k=1}^n F_k^+F_k^-}\;|\t\ra.
\label{fi+Fi+}
\end{equation}
Equation~(\ref{fi+Fi+}) holds for any $|\t \ra$. Therefore
\begin{equation}
f_i^+ =e_{i0}= F_i^+
\sqrt{p-\sum_{k=1}^n F_k^+F_k^-},\qquad i=1,\ldots ,n.
\label{Fermi+}
\end{equation}
In a similar way one derives from~(\ref{fi-})
\begin{equation}
f_i^- =e_{0i}=
\sqrt{p-\sum_{k=1}^n F_k^+F_k^-}\;F_i^-,\qquad i=1,\ldots ,n.
\label{Fermi-}
\end{equation}
Evidently (see~{\ref{N0}}),
 \begin{equation}
e_{00}=
p-\sum_{k=1}^n F_k^+F_k^-,
\end{equation}
and simple calculations lead to
\begin{equation}
e_{ij}=
F_j^+F_i^-, \qquad i,j=1,\ldots,n.
\end{equation}
In such a way we have expressed all Weyl generators
$\{ e_{ij}|i,j=0,1,\ldots ,n \}$ of
$gl(1|n)$ via $n$ pairs of Fermi operators. Let us mention that
the ``ferminization''~(\ref{Fermi+})-(\ref{Fermi-}) of the
Jacobson generators of $sl(1|n)$ is not new. It is known as
the Holstein-Primakoff realization of $sl(1|n)$~\cite{HP}.
The use of boson or fermion operators to construct representations 
of Lie algebras or Lie superalgebras has a long history. 
For Lie superalgebras, see~\cite{{LSR},{Bars}} for a 
realization of the algebra in terms of Bose and Fermi operators 
and a construction of the  corresponding representations. 

Finally, we wish to mention that the Jacobson generators of $gl(1|n)$ and the
considered Fock representations $W(p,n)$ (or equivalently, the quasi-Fermi operators)
are implicitly present in certain physical models.

Examples from condensed matter physics include mainly models related in one or another
way to high-temperature superconductivity. We have in mind those lattice models of
strongly correlated electron systems where (the electronic part of) the Hamiltonian is
expressed in terms of Hubbard operators ($X$-operators)~\cite{Hubbard} as for instance
in \cite{Ruck} or in \cite{Zeyher1}. In such models, each Hubbard operator
is labeled by three indices, $X_A^{ij}$, where $A$ refers to the lattice
site and $i,j=0,1,\ldots,N$, if the (combined spin-flavor) degrees of freedom (the
number of the orbitals) of the electrons at each fixed site are $N$. For any site the
operators $X_A^{i0}$ and $X_A^{0i}$, $i=1,\ldots, N$, are said to be fermion-like 
generators (or of odd degree), whereas $X_A^{00}$, $X_A^{ij}$, $i,j=1,\ldots, N$ are
boson-like generators (or of even degree). The $X$-operators obey the relations
 \beq
 [X_A^{ij},X_B^{kl}]_{\pm}=\delta_{AB}(\delta_{jk}X_A^{il} \pm
 \delta_{il}X_A^{kj}),\quad i,j,k,l=0,\ldots,N, \label{X}
 \eeq
 with the upper signs if both $X$-operators in the left hand side are
 fermion-like and with lower signs  in all other cases. Clearly (\ref{X}) can 
 also be written as
\begin{equation}
\lb X_A^{ij}, X_B^{kl} \rb  = \delta_{AB}(\delta_{jk} X_A^{il} - (-1)^{\deg(X_A^{ij})
\deg(X_A^{kl})} \delta_{il} X_A^{kj}), \label{XLS}
\end{equation}
which indicates,  that  for a fixed value of the lattice
site $A$ the Hubbard operators are the Weyl generators of $gl(1|N)\equiv
gl(1|N)_A$~\cite{Coleman}, cfr.~(\ref{Weyl}). Since moreover for $A\ne B$ the 
$X$-operators supercommute,
$\lb X_A^{ij}, X_B^{kl} \rb =0$, the conclusion is that all Hubbard operators
constitute a basis in the algebra ${\cal L}$ which is a direct sum of all $gl(1|N)_A$,
i.e.,
 \beq
{\cal L}= \bigoplus_A gl(1|N)_A. \label{L}
 \eeq

Each local state space per site $A$ (we suppress the site index $A$ whenever possible)
has a basis consisting of all vectors
\begin{equation}
| n_0;n_1, \ldots,n_N\rangle, \quad n_1,\ldots,n_N\in \{0,1\},
\label{n00}
\end{equation}
 subject to the additional constraint $n_0=p-\sum_{k=1}^N n_k,$ with $p$ fixed
(in~\cite{Zeyher1}, $p=N/2$; and in~\cite{Ruck}, $p=1$).
Since $n_0$ is required to be a non-negative integer, only such sets of numbers $n_1,
\ldots, n_N$, are admitted for which
\begin{equation}
n_1+\ldots+n_N\le p.\label{n0}
\end{equation}
The physical interpretation of the state (\ref{n00}) is that it corresponds to a
configuration with $n_1$ electrons on the first orbital, $n_2$ electrons on the second
orbital, etc.  The action of the
$X$-operators on the states (\ref{n00}) reads (throughout below
$i,j,k=1,\ldots,N$)~\cite{Zeyher1}~:
\begin{eqnarray}
&& X_A^{kk} | n_0;n_1, \ldots,n_k,\ldots,n_N\rangle = n_k | n_0; n_1,
\ldots,n_k,\ldots,n_N\rangle,\label{Xkk} \\
&& X_A^{jk} | n_0;n_1, \ldots,n_j,\ldots,n_k,\ldots,n_N\rangle = \nn\\
&& \qquad  (-1)^{n_j+\cdots+n_{k-1}} | n_0; n_1, \ldots,n_j+1,\ldots,
 n_k-1,\ldots,n_N\rangle,~~~j\ne k, \label{Xjk} \\
&& X_A^{0k} | n_0;n_1, \ldots,n_k,\ldots,n_N\rangle = \nn\\
&& \qquad  \sqrt{n_0+1}(-1)^{n_1+\cdots+n_{k-1}} | n_0+1; n_1, \ldots,
 n_k-1,\ldots,n_N\rangle, \label{X0k} \\
&& X_A^{k0} | n_0;n_1, \ldots,n_k,\ldots,n_N\rangle = \nn\\
&& \qquad  \sqrt{n_0}(-1)^{n_1+\cdots+n_{k-1}} | n_0-1; n_1, \ldots,
 n_k+1,\ldots,n_N\rangle. \label{Xk0}
\end{eqnarray}

\noindent In eqs.\ (\ref{Xkk})-(\ref{Xk0}), the convention is that the vectors on the
right hand sides with unacceptable arguments should be identified with zero.

Equations (\ref{X0k}) and (\ref{Xk0}) clearly indicate that the operator $X_A^{k0}$
(resp. $X_A^{0k}$) creates (resp. annihilates) an electron on the $k$-th orbital of site
$A$. However, these operators are not Fermi creation and annihilation operators in the
strict sense because $\{X_A^{0i},X_A^{j0}\} \ne \delta_{ij}$.

Comparison with the formulas from Section~2 now leads to the
identification of the Hubbard model Hilbert space with the
representation $W(p,n)$, where
\begin{equation}
N=n,\qquad (n_1,\ldots,n_N)=(\t_1,\ldots,\t_n), \qquad n_0=p-|\t|.
\end{equation}
Furthermore, the Hubbard operators (on a fixed site $A$) are expressed in terms of the
Jacobson creation and annihilation operators $f_A^\pm$ of $gl(1|n)$~:
\begin{equation}
X_A^{0k}= f_k^-,\qquad X_A^{k0}=f_k^+, \qquad X_A^{jk}=\{f_j^+,f_k^-\}, \ (j<k).
\end{equation}

So the conclusion is that the operators $X_A^{k0}$ and $X_A^{0k}$, 
creating and annihilating electrons at site
$A$,  are not Fermi operators. They are creation and
annihilation operators of the Lie superalgebra $gl(1|n)$. The statistics of the
electrons or of any other (quasi)particles described with these operators is not 
Fermi statistics, it is $A$-superstatistics of order $p$.

In order to quote an example from
nuclear physics,  note that the $p=1$ Jacobson generators
$f_1^\pm,\ldots,f_n^\pm$ of $sl(1|n)$ together with $N_0$ satisfy the relations~:
\begin{eqnarray}
&&  f_j^- f_i^-=f_i^+ f_j^+ =0 ,\\ \label{NP1}
&&  f_i^- f_j^+ = \delta_{ij}N_0 ,\\
&&  N_0 f_i^+ = f_i^- N_0=0,\\
&&  N_0^2=N_0.
\end{eqnarray}
In  nuclear shell model theory the operators with the above properties are called
{\it ideal odd-particle} (IOP) creation and annihilation operators~\cite{KM1}.
Okubo~\cite{Okubo} refers to the algebra of IOP operators as to the Marshalek algebra. The
IOP operators play a relevant  role for the description of  properties of  odd nuclei
in the frame of the nuclear shell model (see the review article~\cite{KM2} and 
references therein).

\section{Macroscopic properties of $A$-superstatistics}
\setcounter{equation}{0}

To describe the macroscopic properties of $A$-superstatistics
for the Hamiltonian $H$ with orbitals~$i$ ($i=1,\ldots,n$), see (\ref{H}),
it is for us irrelevant whether the different orbitals
correspond to different particles, to different energy levels of
particles, or to different internal states of the particles.
The only assumption is that they satisfy the Pauli principle
for $A$-superstatistics, which follows from the Fock space
construction.

As usually, we assume that the system is in a thermal and diffusive
contact and in a thermal and diffusive equilibrium with a much
bigger reservoir. We denote by $\tau$ its (fundamental) temperature,
by $\mu_i$ the chemical potential and by $\epsilon_i$ the energy
for the particles on orbital~$i$.

The probability ${\cal P}(p,n;\t)$ for the system to be in
a (quantum) state $\t=(\t_1,\ldots,\t_n)$ with
$|\t|=\t_1+\cdots+\t_n$ particles and energy $E=\t_1\epsilon_1+
\cdots+\t_n\epsilon_n$ is given by the expression
\begin{equation}
{\cal P}(p,n;\t) = {1\over Z(p,n)} \exp\left(\sum_{i=1}^n
{\mu_i-\epsilon_i \over \tau}\t_i \right).
\end{equation}
The numerator in this expression is the Gibbs factor of the
system in the state $\t$, and $Z(p,n)$ is the grand partition
function (GPF) of the system. In the case of Bose or Fermi statistics or 
their generalizations (Green parastatistics~\cite{Green}), the GPF  
is simply a product of the GPFs of all orbitals.
This is due to the fact that for those statistics the different orbitals can be
considered as independent subsystems~: the filling of each orbital is completely
independent of the number of particles that have already been accommodated on the other
orbitals. Here due to the new Pauli principle this is no longer the case if $p<n$.
Therefore we have to compute directly the GPF for the whole system. The latter as usual
is the sum of the Gibbs factors over
all possible states of the system. So we have~:
\begin{equation}
Z(p,n)= \sum_{0\leq \t_1+\cdots+\t_n \leq p \atop \t_i\in\{0,1\}}
\left( \exp({\mu_1-\epsilon_1 \over \tau})\right)^{\t_1} \cdots
\left( \exp({\mu_n-\epsilon_n \over \tau})\right)^{\t_n} .
\end{equation}
In terms of the notation
\begin{equation}
x_i=\exp({\mu_i-\epsilon_i \over \tau}),\qquad i=1,\ldots,n,
\label{xi}
\end{equation}
we have
\begin{equation}
Z(p,n) = \sum_{0\leq \t_1+\cdots+\t_n \leq p \atop \t_i\in\{0,1\}}
x_1^{\t_1} x_2^{\t_2}\cdots x_n^{\t_n}
= \sum_{k=0}^{\min(p,n)}
\sum_{ \t_1+\cdots+\t_n=k \atop \t_i\in\{0,1\}}
x_1^{\t_1} x_2^{\t_2}\cdots x_n^{\t_n} .
\end{equation}
It follows that $Z(p,n)=Z(n,n)$ if $p>n$; therefore, we shall from
now on assume that $p\leq n$, thus covering all possible cases.
Since all macroscopic properties are encapsulated in the
grand partition function, this observation also implies that
{\em $A$-superstatistics with $p\geq n$ has the same macroscopic
properties as ordinary Fermi statistics}. So the case $p=n-1$
can be considered as the smallest deviation from Fermi statistics,
as far as the macroscopic properties are concerned.
In the following, we shall sometimes pay special attention to
the $p=n-1$ case, and compare it with the properties of Fermi statistics.

In the present context, it is appropriate to introduce the elementary
symmetric functions $e_k(x_1,\ldots,x_n)$, $k=0,1,\ldots$.
The $k$-th elementary symmetric function~\cite{Macdonald} is the sum over all
products of $k$ distinct variables $x_i$, so that $e_0(x_1,\ldots,x_n)=1$
and
\begin{eqnarray}
e_k(x_1,\ldots,x_n) &=& \sum_{i_1<i_2<\cdots<i_k} x_{i_1}x_{i_2}\cdots x_{i_k} \\
&=& \sum_{ \t_1+\cdots+\t_n=k \atop \t_i\in\{0,1\} }
x_1^{\t_1} x_2^{\t_2}\cdots x_n^{\t_n}.
\end{eqnarray}
For instance, $e_1(x_1,x_2,x_3)=x_1+x_2+x_3$,
$e_2(x_1,x_2,x_3)=x_1x_2+x_1x_3+x_2x_3$,
$e_3(x_1,x_2,x_3)=x_1x_2x_3$, and $e_k(x_1,x_2,x_3)=0$ for $k>3$.
The generating function for the $e_k$ is given by~\cite{Macdonald}
\begin{equation}
\sum_{k=0}^n e_k(x_1,\ldots,x_n) t^k = (1+x_1t)\cdots(1+x_nt).
\label{gf}
\end{equation}
In terms of the elementary symmetric functions, one finds
\begin{equation}
Z(p,n)= \sum_{k=0}^p e_k(x_1,\ldots,x_n).
\label{Z}
\end{equation}
This sum does not simplify if $p<n$; for $p=n$, (\ref{gf}) yields
\begin{equation}
Z(n,n)=(1+x_1)(1+x_2)\cdots(1+x_n).
\end{equation}
Also from (\ref{gf}), it follows that one can give the following
description~: for $p<n$, $Z(p,n)$ consists of those terms of
$Z(n,n)$ that have total degree less than or equal to~$p$.

Let us now consider some other thermodynamic quantities.
The probability ${\cal P}(p,n;\t)$ for the system to be in the
state $\t=(\t_1,\ldots,\t_n)$ with $|\t|$ particles reads
\begin{equation}
{\cal P}(p,n;\t) = {x_1^{\t_1}x_2^{\t_2}\cdots x_n^{\t_n} \over Z(p,n)}.
\end{equation}
Therefore, the average number of particles in the system is
\begin{equation}
\bar N(p,n) = \sum_{0\leq\t_1+\cdots+\t_n \leq p \atop \t_i\in\{0,1\}}
|\t| {\cal P}(p,n;\t)
= \sum_{0\leq \t_1+\cdots+\t_n \leq p \atop \t_i\in\{0,1\}} |\t|
{x_1^{\t_1}x_2^{\t_2}\cdots x_n^{\t_n} \over Z(p,n)}.
\label{N}
\end{equation}
This can be rewritten as
\begin{equation}
\bar N(p,n) = \sum_{i=1}^n x_i \partial_{x_i} \ln (Z(p,n)).
\end{equation}
In terms of symmetric functions, the numerator of~(\ref{N}) reads
\begin{equation}
\sum_{0\leq \t_1+\cdots+\t_n \leq p \atop \t_i\in\{0,1\}} |\t|
x_1^{\t_1}x_2^{\t_2}\cdots x_n^{\t_n} =
\sum_{k=0}^p k
\sum_{ \t_1+\cdots+\t_n=k \atop \t_i\in\{0,1\} }
x_1^{\t_1} x_2^{\t_2}\cdots x_n^{\t_n}
 = \sum_{k=0}^p k e_k(x_1,\ldots,x_n),
\end{equation}
so we can write
\begin{equation}
\bar N(p,n) = { \sum_{k=0}^p k e_k(x_1,\ldots,x_n) \over
\sum_{k=0}^p e_k(x_1,\ldots,x_n) },
\label{N1}
\end{equation}
or equivalently,
\begin{equation}
\bar N(p,n) = p-{ \sum_{k=0}^{p-1} (p-k) e_k(x_1,\ldots,x_n) \over
\sum_{k=0}^p e_k(x_1,\ldots,x_n) }.
\label{N2}
\end{equation}
{}From this last equation it is clear that the average number of
particles in the system is indeed less than $p$.
Formula~(\ref{N1}) simplifies when $p=n$. Indeed, from (\ref{gf})
we have
\begin{eqnarray}
&&\sum_{k=0}^n k e_k(x_1,\ldots,x_n) t^k =
t{\partial\over \partial t}\left( \sum_{k=0}^n e_k(x_1,\ldots,x_n)
t^k\right) \nonumber\\
&& =
t{\partial\over \partial t}\left( \prod_{i=1}^n(1+x_it)\right)=
t \sum_{r=1}^n {x_r\over 1+x_rt} \prod_{i=1}^n(1+x_it),
\end{eqnarray}
and so
\begin{equation}
\bar N(n,n) = \sum_{r=1}^n {x_r\over 1+x_r}.
\label{Ntyp}
\end{equation}

Next, we shall determine the equilibrium distribution of the particles
on the orbitals. First of all, consider the orbital~$n$. Either there
are no particles on this orbital, or else there is just one
particle. Denote by ${\cal P}(p,n;\t_n=0)$, resp.\ ${\cal P}(p,n;\t_n=1)$,
the probability that there are no particles present (resp.\ that there is
one particle present) on orbital~$n$. From the sum of the
corresponding Gibbs factors, one finds~:
\begin{eqnarray}
{\cal P}(p,n;\t_n=0) &=&
{Z(p,n)\vert_{x_n=0} \over Z(p,n) }
= { \sum_{k=0}^p e_k(x_1,\ldots,x_{n-1}) \over
 \sum_{k=0}^p e_k(x_1,\ldots,x_{n-1},x_n)}, \\
{\cal P}(p,n;\t_n=1) &=&
1- {\cal P}(p,n;\t_n=0) =
x_n  { \sum_{k=0}^{p-1} e_k(x_1,\ldots,x_{n-1}) \over
 \sum_{k=0}^p e_k(x_1,\ldots,x_{n-1},x_n)}.
\end{eqnarray}
The last relation follows from the trivial observation
\begin{equation}
e_k(x_1,\ldots,x_{n-1},x_n)=e_k(x_1,\ldots,x_{n-1})+x_n
e_{k-1}(x_1,\ldots,x_{n-1}) .
\label{trivial}
\end{equation}
By the symmetry (of the symmetric functions), these probabilities
extend to any orbital~$i$~:
\begin{eqnarray}
{\cal P}(p,n;\t_i=0) &=&
{Z(p,n)\vert_{x_i=0} \over Z(p,n) }
= { \sum_{k=0}^p e_k(x_1,\ldots,\widehat x_i,\ldots,x_{n}) \over
 \sum_{k=0}^p e_k(x_1,\ldots,x_n)}, \\
{\cal P}(p,n;\t_i=1) &=&
1- {\cal P}(p,n;\t_i=0) =
x_i  { \sum_{k=0}^{p-1} e_k(x_1,\ldots,\widehat x_i,\ldots,x_{n}) \over
 \sum_{k=0}^p e_k(x_1,\ldots,x_n)}.
\end{eqnarray}
Herein, $(x_1,\ldots,\widehat x_i,\ldots,x_{n})$ stands for the
$(n-1)$-tuple obtained by removing $x_i$ from the $n$-tuple
$(x_1,\ldots,x_n)$.
It is now clear that the average number of particles on the $i$-th orbital,
denoted by $\bar \t_i$, is just ${\cal P}(p,n;\t_i=1)$.
In other words~:
\begin{equation}
\bar \t_i = x_i  { \sum_{k=0}^{p-1} e_k(x_1,\ldots,\widehat x_i,\ldots,x_{n}) \over
 \sum_{k=0}^p e_k(x_1,\ldots,x_n)}.
\label{barti}
\end{equation}
{}From~(\ref{Z}) and~(\ref{trivial}) this is also~:
\begin{equation}
\bar \t_i = x_i \partial_{x_i} (\ln Z(p,n) ).
\end{equation}
For $p\geq n$~(\ref{barti}) gives the Fermi case
\begin{equation}
\bar \t_i^{f} = {x_i \over 1+x_i},
\label{bartityp}
\end{equation}
which is consistent with~(\ref{Ntyp}). So $\bar \t_i^{f}$ denotes
the average number of particles on orbital~$i$ in the case of
Fermi statistics. It is interesting to consider the
deviation when $p=n-1$. We can express the average number
of particles on orbital~$i$ in the case of $A$-superstatistics
of order $p=n-1$ by means of the Fermi averages $\bar \t_i^{f}$~:
\begin{equation}
\bar \t_i^{p=n-1} = {\bar \t_i^{f}-\prod_{j=1}^n \bar \t_j^{f} \over
1-\prod_{j=1}^n \bar \t_j^{f}}.
\end{equation}
Clearly, these new averages are small deviations from the averages
in the case of Fermi statistics. Also note that the deviation of the
average on orbital~$i$ depends on the Fermi averages on all other
orbitals.
Similarly the average number of particles in the system for $p=n-1$ is
\begin{equation}
\bar N (p=n-1,n) = {\bar N(n,n)-n\prod_{j=1}^n \bar \t_j^{f} \over
1-\prod_{j=1}^n \bar \t_j^{f}},
\end{equation}
where $\bar N(n,n)$ is the average particle number in the case
of Fermi statistics.

The average energy of the particles on the $i$-th orbital
is given by
\begin{equation}
\bar E_i = \epsilon_i \bar \t_i =\epsilon_i x_i \partial_{x_i} (\ln Z(p,n) ),
\end{equation}
and the average energy of the total system is
\begin{eqnarray}
\bar E(p,n) &=& \sum_{i=1}^n \epsilon_i \bar \t_i
= \sum_{i=1}^n \epsilon_i x_i \partial_{x_i} (\ln Z(p,n) ) \nn\\
&=& {1\over Z(p,n)} \sum_{i=1}^n \epsilon_i x_i
\sum_{k=0}^{p-1} e_k(x_1,\ldots,\widehat x_i,\ldots,x_{n}) .
\end{eqnarray}
We can again express the average energy in the case $p=n-1$
via the average energy in the Fermi case $\bar E(n,n)$, and the
Fermi averages $\bar \t_i^{f}$~:
\begin{equation}
\bar E(p=n-1,n) = {\bar E(n,n)-\sum_{k=1}^n \epsilon_k\prod_{j=1}^n
\bar \t_j^{f} \over 1-\prod_{j=1}^n \bar \t_j^{f}}.
\end{equation}

Also other thermodynamical functions, such as the entropy $S(p,n)$
and the heat capacity $C_V(p,n)$, defined in terms of the grand
partition function and the average energy, can be computed.
The formulation of these expressions in terms of symmetric
functions does not lead to further simplifications or insights,
so we shall not deal with these.

\section{$A$-superstatistics in the degenerate case}
\setcounter{equation}{0}

Let us consider as a particular example a Hamiltonian of the form
\begin{equation}
H=\epsilon \sum_{i=1}^n N_i.
\end{equation}
Thus we assume that all orbitals have the same energy, and let
us furthermore assume that they also have the same chemical
potential, i.e.\ $\mu_1=\mu_2=\cdots=\mu_n=\mu$. Therefore 
$x_1=x_2=\cdots =x_n =x$, with
\begin{equation}
x=\exp ({\mu-\epsilon \over\tau}).
\end{equation}
In this case the orbitals label internal degrees of freedom of 
the particles such as spin, color, flavor, etc.
The thermodynamical functions for this example follow from the
formulas of the previous section, under the specialization
$x_i=x$ ($i=1,\ldots,n$).

Since the number of terms in $e_k(x_1,\ldots,x_n)$ is given by
${n \choose k}$, we have
\begin{equation}
e_k(\underbrace{x,\ldots,x}_{\hbox{$n$ times}}) = {n \choose k} x^k.
\end{equation}
Thus~(\ref{Z}) yields
\begin{equation}
Z(p,n)= \sum_{k=0}^p {n \choose k} x^k.
\label{Zx}
\end{equation}
For $p<n$ this cannot be rewritten in a closed form; for $p\geq n$,
this is simply $(1+x)^n$, i.e.\ the GPF for a Fermi system with $n$
distinct orbitals having the same energy.

For $p<n$ the sum in~(\ref{Zx}) can be rewritten as follows~:
\begin{eqnarray}
Z(p,n) &=& \sum_{k=0}^n {n \choose k} x^k - \sum_{k=p+1}^n {n \choose k} x^k\nn\\
&=&(1+x)^n-{n \choose p+1} x^{p+1}\;{}_2F_1\left({1,p-n+1 \atop p+2};-x\right),
\label{Zx1}
\end{eqnarray}
where ${}_2F_1$ is the classical hypergeometric series
\beq
{}_2F_1\left({a,b \atop c};x\right) = \sum_{k=0}^\infty {(a)_k(b)_k \over{(c)_k}}
{x^k \over{k!}}, ~~~(d)_k=d(d+1)\ldots (d+k-1).
\eeq
The first term in the right hand side of~(\ref{Zx1})
is the Fermi GPF and therefore the second term gives the
difference between Fermi statistics and $A$-superstatistics.
In a sense, it describes the statistical interaction between the particles.

Using Euler's transformation formula for hypergeometric
functions, i.e.
\beq
{}_2F_1\left({a,b \atop c};x\right) = (1-x)^{c-a-b}
{}_2F_1\left({c-a,c-b \atop c};x\right),
\eeq
(\ref{Zx1}) can also be rewritten as
\beq
Z(p,n)= (1+x)^n \left( 1- {n\choose p+1} x^{p+1}\;
{}_2F_1\left({p+1,n+1 \atop p+2};-x\right)\right).
\label{Zx2}
\eeq
So (\ref{Zx1}) gives the deviation from the Fermi GPF in
additive form, and (\ref{Zx2}) gives the deviation in multiplicative
form.

The average number of particles follows from~(\ref{N1})~:
\begin{equation}
\bar N(p,n) = {\sum_{k=0}^p k {n\choose k} x^k \over
\sum_{k=0}^p {n\choose k} x^k}.
\label{N3}
\end{equation}
Using the definition of hypergeometric functions, equation~(\ref{N3}) can be
rewritten as
\beq
\bar N(p,n)={nx (1+x)^{n-1}-(p+1)  {n\choose p+1} x^{p+1}\;
{}_2F_1\left({1,p-n+1 \atop p+1};-x\right)\over
(1+x)^{n}-  {n\choose p+1} x^{p+1}\;
{}_2F_1\left({1,p-n+1 \atop p+2};-x\right)}.
\eeq
For $p\geq n$, this becomes
\begin{equation}
\bar N(p\geq n ,n) = {nx \over 1+x}.
\end{equation}

The average number of particles on the $i$-th orbital follows
from~(\ref{barti})~:
\begin{equation}
\bar\t_i = x { \sum_{k=0}^{p-1} {n-1\choose k} x^k \over
\sum_{k=0}^p {n\choose k} x^k}.
\end{equation}
For $p\geq n$, this sum simplifies to $x\over 1+x$. In the general
case, simple properties of binomial coefficients lead to~:
\begin{equation}
\bar\t_i = {x\over 1+x} -{n-1\choose p} {x^{p+1} \over (1+x) Z(p,n)}.
\end{equation}
Obviously, this expression is the same for every orbital~$i$.
As a consequence, the average number of particles of the total
system can be rewritten as
\begin{equation}
\bar N(p,n) = {nx \over 1+x}
-n{n-1\choose p} {x^{p+1} \over (1+x) Z(p,n)}.
\end{equation}

It is interesting to consider an example. Let $n=5$; we shall
examine the dependence of the average number of particles on the $i$-th 
orbital $\bar\t_i$ upon the variable
\begin{equation}
y={\epsilon-\mu \over \tau}, \hbox{ where } x=e^{-y}.
\end{equation}
In Figure~1, we plot $\bar\t_i$
for $p=1,2,\ldots,5$. The case $p=n=5$ yields the known Fermi-Dirac
distribution function. For $p>n$, the distribution function is the same.
For $p<n=5$, the distribution function is different. The difference
is most noticable for $\epsilon < \mu$ (or $y<0$), as the average number 
of particles cannot exceed $p/n$. 

The case $p=1$ and any $n$ is also of interest~:
\beq
\bar N(1,n)={n\over{e^{(\epsilon -\mu)/ \tau}}+n}.
\label{4.17}
\eeq
$\bar N(1,n)$ is always smaller than $1$, i.e.\
the system can accommodate at most one particle.
When $n=1$ this corresponds to the Fermi-Dirac distribution function.
When $n$ increases the average number of particles of the system
increases also for fixed $y={\epsilon -\mu \over \tau}$.
This description ($p=1$, $n>1$) corresponds
to a system consisting of ``hard-core bosons''.
Such particles appear (as mentioned in Sect. 3) in some models of condensed matter
physics~\cite{Ruck} and nuclear physics~\cite{KM2}.
This case is illustrated in Figure~2, where we take
$p=1$ fixed and let $n$ vary.

\section{Equidistant energy levels}
\setcounter{equation}{0}

Now we consider the Hamiltonian~(\ref{H})
with equidistant energies $\epsilon_i$.
Let the gap between the different energy
levels be $\Delta >0$.
Then
\begin{equation}
\epsilon_i=\epsilon_1+(i-1)\Delta, \qquad (i=1,2,\ldots,n).
\label{eni}
\end{equation}
We  assume also
that $\mu_1=\mu_2=\cdots=\mu_n=\mu$.
Under these conditions the different orbitals correspond to different
energy levels.
According to notation of~(\ref{xi}), we have
\begin{equation}
x_i= \exp \left( {\mu-\epsilon_i\over \tau} \right)
= \exp \left( {\mu-\epsilon_1\over \tau} \right)
\exp \left(- {\Delta\over \tau} \right)^{i-1}
= x q^{i-1},
\label{xiequid}
\end{equation}
where we have used the notation
\begin{equation}
x=x_1=\exp \left( {\mu-\epsilon_1\over \tau} \right)
\qquad\hbox{ and }\qquad
q=\exp \left( -{\Delta\over \tau}\right).
\label{xq}
\end{equation}
Under this specialization the elementary symmetric functions
simplify. For this purpose, consider their generating
function~(\ref{gf}). Using~\cite[p.~26]{Macdonald} one finds,
\bea
&&(1+xt)(1+qxt)\cdots(1+q^{n-1}xt)=\nn \\
&& \sum_{k=0}^n q^{k(k-1)/2} \left[ n \atop k \right] x^k t^k=
\sum_{k=0}^n q^{k(k-1)/2} { (q^{n-k+1};q)_k \over (q;q)_k} x^kt^k,
\eea
where $\left[ n \atop k \right]$ denotes the Gaussian
polynomial~\cite[p.~26]{Macdonald} and
$(a;q)_k$ the $q$-raising factorials~\cite{GR}
\beq
\left[ n \atop k \right] =
{(1-q^n)(1-q^{n-1})\cdots (1-q^{n-k+1}) \over
(1-q)(1-q^2) \cdots (1-q^k)},
\label{5.5}
\eeq
\beq
(a;q)_k = (1-a)(1-qa)\cdots (1-q^{k-1}a).
\label{5.6}
\eeq
It follows from~(\ref{gf}) and (\ref{xiequid}) that
\beq
e_k(x,qx,q^2x,\cdots,q^{n-1}x)= q^{k(k-1)/2}\left[ n \atop k \right] x^k=
q^{k(k-1)/2}{ (q^{n-k+1};q)_k \over (q;q)_k}x^k.
\label{5.7}
\eeq

To write down the GPF, we use~(\ref{Z}), the specialization~(\ref{xiequid})
and~(\ref{5.7}) 
\beq Z(p,n)= \sum_{k=0}^p q^{k(k-1)/2}\left[ n \atop k \right] x^k=
\sum_{k=0}^p q^{k(k-1)/2}{ (q^{n-k+1};q)_k \over (q;q)_k}x^k. \label{5.8} 
\eeq 
Using the formula 
\beq \left[ n \atop k \right] = q^{-k(k-1)/2}(-q^n)^k{ (q^{-n};q)_k \over
(q;q)_k}, \label{5.81} 
\eeq 
one obtains 
\beq Z(p,n)= \sum_{k=0}^p { (q^{-n};q)_k \over
(q;q)_k}(-q^n)^kx^k= {\ }_2\phi_1\left( {q^{-n}, q^{-p} \atop q^{-p} } ;-q^nx \right).
\label{5.82} 
\eeq  
The last function is a terminating basic hypergeometric
series~\cite{GR},
\beq 
{\ }_2\phi_1\left( {a, b \atop c } ;x \right)
= \sum_{k=0}^\infty { (a;q)_k (b;q)_k\over
(c;q)_k}{x^k\over{(q;q)_k}} .
\eeq 
For $p=n$~(\ref{5.8}) yields \beq Z(n,n)=  (-x;q)_n \label{5.83} \eeq

The average number of particles in the system follows from~(\ref{N1})~:
\beq
\bar N(p,n) = { \sum_{k=0}^p k q^{k(k-1)/2} \left[ n \atop k\right] x^k \over
\sum_{k=0}^p q^{k(k-1)/2}\left[ n \atop k\right] x^k } = x\ {\partial\over
\partial x} ( \ln Z(p,n) ) \quad \Big(= \tau\ {\partial\over \partial
\mu}( \ln Z(p,n) )\Big). \label{5.9}
\eeq
For $p=n$~(\ref{5.9}) becomes
\beq
\bar N(n,n) = x \sum_{i=0}^{n-1} {q^i  \over 1+q^i x}.
\label{5.91}
\eeq

The average number of particles on the $i$th orbital can be written in the form~:
\beq
\bar \theta_i = {xq^{i-1}\over Z(p,n)}\sum_{k=0}^{p-1}  \sum_{l=0}^{k} (-1)^{l}
q^{l(i-1)+(k-l)(k-l-1)/2}
\left[ n \atop k-l \right] x^k.
\label{5.10}
\eeq
We used that~:
\beq
e_k(x,qx,q^2x,\ldots,q^{i-2}x,q^ix,\ldots,q^{n-1}x) =
  \sum_{l=0}^{k} (-1)^{l}
q^{l(i-1)+(k-l)(k-l-1)/2}
\left[ n \atop k-l \right] x^k.
\label{5.15}
\eeq
The conclusion from~(\ref{5.10}) is that the ``population'' of the
orbitals depends mainly on their level $i$ via $q^{i-1}$ with
$q=\exp (-\Delta /\tau)<1$~: as $i$ increases,
$\bar \theta_i$ decreases.

Consider the case $p=1$ and any $n$. Then
\beq
Z(1,n)= 1+(1+q+q^2+\ldots +q^{n-1})e^{(\mu-\epsilon_1)/\tau},
\label{5.16}
\eeq
\beq
\bar N(1,n) = { (1+q+\ldots+q^{n-1}) \over e^{(\epsilon_1-\mu)/\tau}
+ (1+q+\ldots+q^{n-1})}.
\label{5.17}
\eeq
If $q=\exp(-\Delta/\tau)\ll 1$, i.e., for
large gaps between the energy levels or very low temperature,
one can neglect all positive powers of $q$ in~(\ref{5.17}).
What remains is the Fermi-Dirac distribution
\beq
\bar N(1,n) \approx {1\over {e^{(\epsilon_1-\mu)/\tau} +1}}.
\label{5.18}
\eeq
The expression for the average number of particles
on orbital~$i$ reads
\beq
\bar \theta_i = {q^{i-1} \over e^{(\epsilon_1-\mu)/\tau}+(1+q+\ldots+q^{n-1})},
\quad i=1,\l,n. \label{5.19}
\eeq
For very low temperatures, or a large energy gap $\Delta$, (\ref{5.19}) reduces to
\beq
\bar \theta_1 \approx  {1\over
{e^{(\epsilon_1-\mu)/\tau}+1}} \ \hbox{ and }
\ \qquad  \bar \theta_i \approx 0 \
\hbox{ if }\ i>1. \label{5.20}
\eeq
Therefore  if the system
contains a particle, it is ``sitting" permanently on the first
(i.e.\ the lowest) energy orbital.
This also explains why $\bar N(1,n)\approx \bar \theta_1$.
To illustrate these ideas, we plot the values of the average number
of particles $\bar \theta_i$ on orbital~$i$, in the case $p=1$ and
$n=5$ (five equidistant energy levels), as a function of $q$,
see Figure~3.
For any value of $q$ ($0<q<1$), $\bar \theta_1>\bar \theta_2>\bar \theta_3
> \bar \theta_4>\bar \theta_5$.
For small values of $q$, $\bar \theta_1$ is the large and the
other averages close to zero. For increasing values of $q$, the averages
on the other orbitals become larger.

\section{Concluding remarks}

In this paper we have studied the microscopic and thermal (macroscopic) properties
of ``free'' particles, interacting only via statistical interaction. This interaction
follows from the Pauli principle of $A$-superstatistics~: the system cannot
accommodate more than $p$ particles if the order of statistics is $p$, irrespective
of the number of available orbitals (which may even be infinite).

$A$-superstatistics is defined by the triple relations (\ref{2.4})-(\ref{2.6}) 
which should hold for the creation and annihilation operators. The Fock spaces 
for $A$-super\-sta\-tis\-tics are naturally related to certain representations of the
Lie superalgebra $sl(1|n)$, and are labeled by a positive integer $p$ referred to
as the order of statistics.
It is the mathematical structure of the Fock spaces that gives rise to 
the Pauli principle of $A$-superstatistics.

It is shown that the creation and annihilation operators of $A$-superstatistics
are fermion-like. Just like ordinary Fermi operators, the creation (resp.\ annihilation)
operators anti-commute. However, they do not satisfy all traditional Fermi relations~:
only in the limit $p\rightarrow\infty$ do the remaining relations tend to the ordinary
Fermi relations. 
It may be unusual and unconventional to consider such alternative creation and 
annihilation operators. However, we show that (representations of) such operators 
have already appeared in physical models, proving their applicability.

In the second part of the paper we focuss on the macroscopic properties of 
$A$-superstatistics, for a free Hamiltonian. 
The usual thermal functions (grand partition function, average number of
particles, orbital distribution, average energy, $\ldots$) are expressed
in terms of elementary symmetric functions, and can be considered as deviations
of the usual Fermi case when $p<n$.

In addition to the general case, we have considered two specific examples.
The case with identical energy levels per orbital (degenerate case)
leads to further simplifications
of the thermal functions, and the deviation from Fermi statistics (for $p<n$) becomes
more apparent. The corresponding distribution functions are reminiscent of the
Fermi-Dirac case, but there are also some striking differences (see Figures~1 and~2).

The case with equidistant energy levels is interesting from the mathematical point
of view, since the thermal functions have simple expressions in terms of $q$-series
(or basic hypergeometric series). This can be considered as a $q$-deformation of the
degenerate case, and in fact the degenerate case can be deduced from the current one
under the limit $q\rightarrow 1$. Also the situation with equidistant energy levels
yields some interesting physical interpretations, e.g.\ concerning the orbital
distribution (see Figure~3).

In this paper, $A$-superstatistics was introduced mainly on the ground of 
mathematical considerations. 
Therefore, a natural next step would be to use this statistics for studying 
real physical objects or phenomena like an ideal $A$-gas or the quantum Hall 
effect, or statistics of cosmic rays etc.
Such a study also includes a comparison of the obtained
results with existing experimental data or with the predictions of other 
noncanonical quantum statistics (parastatistics, fractional statistics, quon 
statistics, anyons statistics, Tsallis statistics or quantum group noncommutative
statistics). We hope to return to this issue elsewhere.

\section*{Acknowledgements}

NIS has been supported by a Marie Curie Individual Fellowship
of  the European Community Programme ``Improving the Human Research Potential
and the Socio-Economic Knowledge Base" under contract number
HPMF-CT-2002-01571. The work was supported 
by grant $\Phi$-910 of the Bulgarian Foundation for Scientific Research.

\newpage
\begin{figure}[htb]
\[
\includegraphics{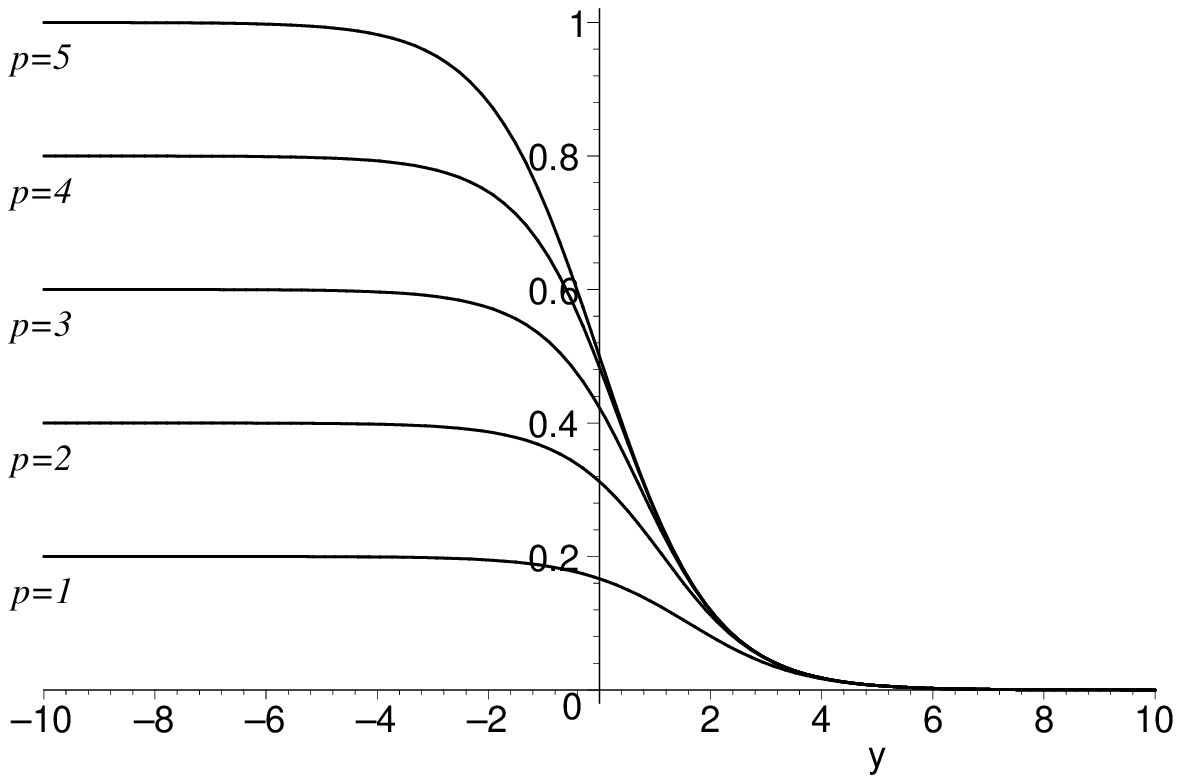}
\]
\caption{Dependence of the average number of particles on the $i$-th orbital 
$\bar\t_i$ upon the variable $y=(\epsilon-\mu)/\tau$, for fixed $n=5$, and $p=1,2,3,4,5$.
The distribution $\bar\t_i$ is independent of $i$ since we consider the degenerate case here.
For $p=n=5$, the distribution function coincides with the Fermi-Dirac
distribution; for $p<n$ it is different and $\bar\t_i$ cannot
exceed $p/n$.}
\end{figure}

\newpage
\begin{figure}[htb]
\[
\includegraphics{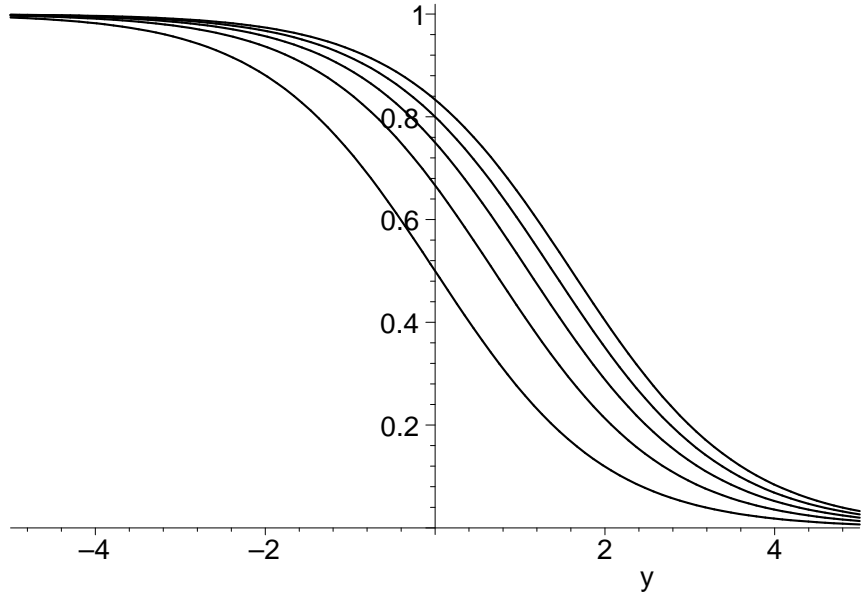}
\]
\caption{Dependence of the average number of particles $\bar N(p,n)$
upon the variable $y=(\epsilon-\mu)/\tau$, for fixed $p=1$, and $n=1,2,3,4,5$,
in the degenerate case.
The graph of $\bar N(1,1)$ is the closest to the $y$-axis, then
$\bar N(1,2)$, etc.
Observe that the graph of $\bar N(1,1)$ coincides with the Fermi-Dirac
distribution.}
\end{figure}

\newpage
\begin{figure}[thb]
\[
\includegraphics{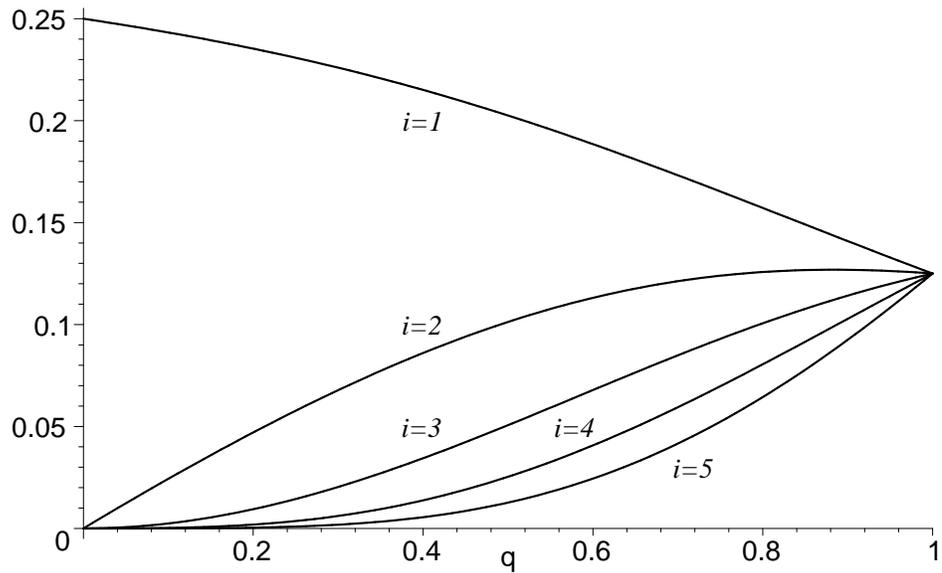} 
\]
\caption{Graphs of the average number of particles $\bar\t_i$ on the
$i$-th orbital, for $p=1$, $n=5$, and $i=1,2,3,4,5$.
The distribution $\bar\t_i$ is plotted as a function of $q=\exp(-\Delta/\tau)$,
where $\Delta$ is the gap between the equidistant energy levels.}
\end{figure}

\end{document}